\documentclass{article}
\usepackage{graphicx}
\usepackage {amsmath}
\usepackage{authblk}
\usepackage{float}
\usepackage{amsthm}
\usepackage{amssymb}
\usepackage[]{blindtext}
\usepackage{hyperref}
\title{Energy Conditions in $f(R,T)$ Gravity  with an Anisotropic Background} 

\author[1]{V. R. PATIL}
\author[2]{J. L. PAWDE}
\author[3]{R. V. MAPARI}
\author[4]{S. K. WAGHMARE}

\affil[1,2]{ Department of Mathematics, Arts, Science \& Commerce College, Chikhaldara, Dist. Amravati, 444 807 (India)}
\affil[3]{Department of Mathematics, Government Science College, Gadchiroli, 442 605 (India)}
\affil[4]{Department of Mathematics, TGPCET, Nagpur, 444807 (India)}

\date{}

\begin{document}

\maketitle
{{\bf Email:} $^3$r.v.mapari@gmail.com}
\begin{abstract}
In our study, we explored the properties of a spatially homogeneous and anisotropic Bianchi type $VI_0$ Universe. Our investigation centered on integrating cosmic domain walls into the $f(R,T)$ theory of gravitation, initially proposed by Harko et al. in 2011. To tackle the field equations, we employed the relationship between the expansion scalar ($\theta$) and the shear scalar ($\sigma$). Our analysis encompassed both the dynamic and cosmological aspects of the Universe. By comparing our findings to the $\Lambda CDM$ model, specifically focusing on the evolution of the jerk parameter, we found a striking agreement between the two models. A noteworthy discovery was the verification of accelerated expansion in our described model, consistent with the prevailing observational data. Finally, we examine the energy condition criteria and determine that the violation of the Strong Energy Condition (SEC), while the Null Energy Condition (NEC), Weak Energy Condition (WEC) and Dominant Energy Condition (DEC) continue to meet the requirements for positivity.
\end{abstract}
{\bf Keywords:}{Bianchi type $VI_0$ cosmological model, $f(R,T)$ gravity, Domain wall, Power law}

\section{Introduction}
In contemporary cosmology, a significant revelation is that the Universe is undergoing both expansion and acceleration. This late-time accelerated expansion has been confirmed through research on high red-shift supernovae (Riess et al.,1998, Perlmutter et al., 1999, Bennet et al., 2003 ). The investigation of the Universe, which seems pervaded by dark energy, has captivated numerous scientists. Research findings have made it evident that the Universe is predominantly influenced by a unique energy form with negative pressure commonly termed dark energy. The cosmological constant, as explored in Padmanabhan's work (Padmanabhan, 2003), plays a pivotal role in deciphering the nature of this dark energy. Evidence from diverse sources like the Cosmic Microwave Background Radiation (CMBR) and supernova surveys has unveiled that the Universe's energy composition comprises about 4\% regular baryonic matter, 22\% dark matter, and 74\% dark energy (Riess et al.,2004, Eisenstein et al., 2005, Astier et al., 2006, Spergel et al., 2007). Recent times have witnessed the proposition of several alterations to the theory of general relativity (GR) to provide a natural gravitational framework for understanding dark energy. In the quest to explain the Universe's late-time acceleration, researchers are actively exploring alternative avenues. Among these, the $f(R)$ theory of gravity stands out as a suitable candidate due to its cosmological implications. Introduced by replacing the Einstein-Hilbert action of GR with a generalized function of the Ricci scalar $R$ (Nojiri and Odintsov, 2007, Multamaki and Vilja, 2006,2007, Shamir, 2010 ), $f(R)$ gravity embodies the amalgamation of early-time inflation and the Universe's late-time acceleration.\\
\paragraph{} Studies on the Bianchi type-$VI_0$ cosmological model indicate its potential convergence towards isotropy (Adhav et al., 2011). Analyzing this model further, it becomes evident that its accelerated expansion is attributed to a negative barotropic equation of state (Yadav et al., 2012). With the passage of time, the deceleration of the Bianchi type-$VI_0$ Universe increases gradually, eventually reaching a constant value (Shaikh and Kaatore, 2016). Scholars (Bali and Kumari, 2017, Satish and Venkateswarlu, 2019) have explored various aspects of this model, including its shearing, non-rotational, and expanding nature. In the context of Lyra geometry, researchers (Pradhan et al., 2007) have investigated solutions for a bulk viscous plane symmetric Universe and highlighted the absence of big bang singularities . Another examination of solutions involving a plane symmetric cosmological model with a domain wall demonstrates the presence of radiation (Pawar et al., 2009). A significant discovery involves solutions for a non-static Bianchi type-III cosmological model with domain walls, both in the presence and absence of a magnetic field, within the framework of general relativity (Adhav et al., 2009). Lastly, in the presence of string and domain walls associated with quarks, researchers (Sahoo and Mishra, 2013) have identified a vacuum kink model that notably lacks a singularity at $r=2k$ . 
\paragraph{} Modifications to General Relativity (GR) involved coupling matter and geometry through a Lagrangian dependent on the stress-energy tensor trace $(T)$ and the Ricci scalar $(R)$. The $f(R,T)$ gravity field equations are derived from the Hilbert-Einstein principle (Harko et al., 2011). Despite setting up energy density and pressure for dark components, equilibrium thermodynamics isn't achievable in $f(R,T)$ gravitation. In thermal equilibrium, photons and non-photons follow the Generalized Second Law of Thermodynamics (Sharif and Zubair, 2012). Analyzing the Kantowski-Sachs bulk viscous Universe showed pressure, density, viscosity, Hubble parameter tend toward zero for high cosmic time (Khade et al., 2018). Investigating different $f(R,T)$ gravity models (with $n=0$ and $n\neq 0$) revealed an expanding, shearing, non-rotating, accelerating Universe (Hasmani and Al- Haysah, 2019). Under dark energy's influence, expansion occurs; negative deceleration and positive Hubble parameter imply exponential expansion and acceleration (Pawar et al., 2019). The imaginary dark energy form acts as quintessence field when $\omega=\frac{1}{3}$, explaining accelerated expansion (Islam et al., 2019). With a stiff fluid, the Bianchi type-$V$ Universe shows isotropy in $f(R,T)$ gravity (Patil et al., 2020). Using domain walls as fractal cosmology's matter source, the flat Friedmann-Robertson-Walker model contracts and accelerates (Pawar et al., 2020). Linear and nonlinear $f(R,T)$ gravity leads to accelerated expansion like $\Lambda$CDM model (Sahoo et al., 2021). Bianchi type models depict shearing, non-rotation, accelerated expansion, approaching isotropy as cosmic time increases (Chaubey and Shukla, 2017).\\
\paragraph{} In the context of $f(R, T)$ gravity, where a perfect fluid serves as the energy source, it has been observed that the geometry of a Bianchi Type $VI_0$ Universe remains undisturbed. However, there is a slight alteration in the matter distribution, as discussed in reference (Rao and Nilima, 2013). For Bianchi Type III and Kantowski-Sachs Universes, predictions indicate a future collapse attributed to domain walls within the $f(R, T)$ theory. This collapse contributes to the model's stability, ensuring the absence of singularities in the Universe (Katore and Hatkar, 2016). The impact of bulk viscosity is noted on pressure and the equation of state parameter, yet it does not affect the density of the domain wall itself (Mahanta et al.,2018). The presence of domain walls leads to an expanding nature of Riemannian space-time, particularly for extended periods. This expansion aligns with observations related to Type Ia supernovae (Shaikh and Wankhade, 2018). In the initial stages, the Universe exhibits expansion with a finite volume that further increases with time. As time ($t$) progresses, the model tends towards isotropy at $t=0$ (Pawar et al.,2021). By delving into the study of a Bianchi type-V Universe within the framework of $f(R, T)$ gravity, and considering the presence of both domain walls and quark matter, it is established that pressure and density experience growth as redshift ($z$) increases. This scenario points towards the existence of a Big Bang singularity and confirms the Universe's accelerated expansion, resembling observations related to Type Ia Supernovae (Maurya et al., 2020).
Moreover, researchers have explored the Plane Symmetry cosmological model within the context of $f(R, T)$ gravity with interacting fields. The outcomes of this study align closely with recent observational data, revealing an expanding Universe (Pawar and Mapari, 2022). \\
The collaborative research endeavors inspire a deeper exploration of the spatially homogeneous anisotropic Bianchi Type $VI_0$ Universe. In this context, we contemplate a Universe replete with a cosmic domain wall that functions as the wellspring of energy within the paradigm of the $f(R, T)$ theory of gravitation.
\section{Basic Equations of $f(R,T)$ Gravity}
The action principle for $f(R,T)$ modified theory is given by 
\begin{equation} \label{e1}
	S=\frac{1}{16\pi G} \int f(R,T)\sqrt{-g}  d^{4} x+\int L_{m} \sqrt{-g}  d^{4} x
\end{equation}
where, the gravitational Lagrangian contains  of an arbitrary function of Ricci scalar $(R)$ and the trace $(T)$ of the energy-momentum tensor $T_{ij}$ of the matter source. $L_m$ denotes matter
Lagrangian density and the $G$ is the gravitational constant considered as $G = 1$.

By modifying the action principle with respect to $g_{ij}$, the associated field equations of $f(R,T)$ gravity are determined as,
\begin{equation} \label{e2} 
	f_{R} \left(R,T\right)R_{ij} -\frac{1}{2} f\left(R,T\right)g_{ij} +\left(g_{ij} \nabla ^{i} \nabla _{i} -\nabla _{i} \nabla _{j} \right)f_{R} \left(R,T\right) =8\pi T_{ij} -f_{T} \left(R,T\right)T_{ij} -f_{T} \left(R,T\right)\Theta _{ij}
\end{equation} 
Where, $f_{R} =\frac{\delta f\left(R,T\right)}{\delta R} $, $f_{T} =\frac{\delta f\left(R,T\right)}{\delta T} $, $\nabla^{i}\nabla_{i}=$D'Alembertian's operator, $T_{ij}$ is the energy momentum tensor and 
\begin{equation} \label{e3}
	\Theta _{ij} =g^{\alpha \beta } \frac{\delta T_{\alpha \beta } }{\delta g^{ij}}
\end{equation}
Here covariant derivative is represented by $\nabla_i$ and the energy momentum tensor $T_{ij}$ emerges from the Lagrangian $L_m$.  By assuming the function $f(R, T) = f(R)$, equation (\ref{e2}) is reduced to field equations of $f(R)$ gravity.
In this article we assume $f(R, T)$ of the form,
\begin{equation} \label{e4} 
	f\left(R,T\right)=f_{1} \left(R\right)+f_{2} \left(T\right) 
\end{equation}
where
\begin{equation} \label{e5} 
	f_{1} \left(R\right)= \lambda R \ \ \  and \ \ \  f_{2} \left(T\right)=\lambda T 
\end{equation}
and $\lambda $ is arbitrary constant.\\
\section{Metric and Field Equations}
The spatially homogeneous and anisotropic Bianchi type $VI_0$ line element can be written in the form as 
\begin{equation} \label{e7} \
	ds^{2}= dt^{2} - A^{2} dx^{2} - B^{2} e^{2x} dy^{2} - C^{2} e^{-2x} dz^{2}  
\end{equation} 
where $A$, $B$ and $C$ are the functions of time $t$ only.\\
The energy momentum tensor for domain wall is
\begin{equation} \label{e8} 
	T_{ij} =\left(g_{ij} +\omega _{i} \omega _{j} \right)\rho -p\omega _{i} \omega _{j}  
\end{equation} 
where $p$ and $\rho $ are the pressure and density of the fluid respectively and  $\omega _{i}$ is four velocity vector satisfying ${\omega }_i{\omega }^j=0$ and  $\omega _{i} \omega ^{i} =-1$.\\
With the help of co-moving coordinates system and from equation (\ref{e3}), (\ref{e4}) the Einstein field equations (\ref{e2}) for the cosmological model (\ref{e7}) are given by
\begin{equation} \label{e9} 
	\frac{\ddot{B} }{B } +\frac{\ddot{C} }{C } +\frac{\dot{B} \dot{C} }{B C } +\frac{1}{A^{2} } =\kappa \rho -{\rm \Lambda } 
\end{equation} 
\begin{equation} \label{e10} 
	\frac{\ddot{A} }{A } +\frac{\ddot{C} }{C } +\frac{\dot{A} \dot{C} }{A C } -\frac{1}{A^{2} } =\kappa \rho -{\rm \Lambda } 
\end{equation}
\begin{equation} \label{e11} 
	\frac{\ddot{A} }{A } +\frac{\ddot{B} }{B } +\frac{\dot{A} \dot{B} }{A B } -\frac{1}{A^{2} } =\kappa \rho -{\rm \Lambda } 
\end{equation} 
\begin{equation} \label{e12} 
	\frac{\dot{A} \dot{B} }{A B } +\frac{\dot{B} \dot{C} }{B C } +\frac{\dot{A} \dot{C} }{A C } -\frac{1}{A^{2} } =-\kappa p-{\rm \Lambda } 
\end{equation} 
\begin{equation} \label{e13} 
	\frac{\dot{B} }{B } -\frac{\dot{C} }{C } =0 
\end{equation}
The overhead dot (.) denotes the derivative with respect to time $t$.\\
Integrating equation (\ref{e13}) we have,
\[B =lC \] 
Where $l$ is a constant of integration.\\
Without loss of generality we choose $l=1$ then we have 
\begin{equation} \label{e14} 
	B =C  
\end{equation}
Using equation (\ref{e14}), equations (\ref{e9})-(\ref{e12}) reduced to
\begin{equation} \label{e15} 
	2\; \frac{\ddot{B} }{B } +\frac{\dot{B}^{2} }{B^{2} } +\frac{1}{A^{2} } =\kappa \rho -{\rm \Lambda } 
\end{equation} 
\begin{equation} \label{e16} 
	\frac{\ddot{A} }{A } +\; \frac{\ddot{B} }{B } +\frac{\dot{A} \dot{B} }{A B } -\frac{1}{A^{2} } =\kappa \rho -{\rm \Lambda } 
\end{equation} 
\begin{equation} \label{e17} 
	2\frac{\dot{A} \dot{B} }{A B } +\frac{\dot{B}^{2}}{B^{2}} -\frac{1}{A^{2} } =-\kappa p-{\rm \Lambda } 
\end{equation}

\section{Solution of Field Equations}
There are three linearly independent equations with five unknowns $A$, $B$, $p$, $\rho$ and $\Lambda$. Therefore to solve this system of equations we assume that the expansion scalar is proportional to shear scalar. This condition leads to 
\begin{equation} \label{e18} 
	A =B^{n} , \ \ \ \ \ \ n\neq0 
\end{equation} 
where, $n$ is positive constant.
From equations (\ref{e15}) and (\ref{e16}) we have
\begin{equation} \label{e19} 
	\frac{\ddot{B} }{B } + \frac{\dot{B}^{2}}{B^{2}} +\frac{2}{A {}^{2} } -\frac{\ddot{A} }{A } -\frac{\dot{A} \dot{B} }{A B } =0 
\end{equation} 
Using equation (\ref{e18}) in (\ref{e19}) we get 
\begin{equation} \label{e20} 
	B =\left[n\left(k_{1} t+k_{2} \right)\right]^{\frac{1}{n} }  
\end{equation} 
Where $k_1$ \& $k_2$ are the constants of integration and \\ $k_{1}^{2} =\frac{1}{n-1} $ ,$n\ne 1$\\
From equation (\ref{e14}), (\ref{e18}) and (\ref{e20}) we have
\begin{equation} \label{e21}
	A = n\left(k_{1} t+k_{2} \right) \  \ \& \  \ C =n\left(k_{1} t+k_{2} \right)^{\frac{1}{n} }
\end{equation}
Using values of $A $, $B $ and $C $, equation (\ref{e7}) yields     
\begin{equation} \label{e22} 
	ds^{2} =dt^{2} -n^{2} \left(k_{1} t+k_{2} \right)^{2} dx^{2} -\left[n\left(k_{1} t+k_{2} \right)\right]^{\frac{2}{n} } \left[e^{2x} dy^{2} -e^{-2x} dz^{2} \right] 
\end{equation} 
The Volume ($V$) is 
\[V=A B C \] 
\begin{equation} \label{e23} 
	V=\left[n\left(k_{1} t+k_{2} \right)\right]^{\frac{n+2}{n} }  
\end{equation} 
The Hubble Parameter ($H$) is
\[H=\frac{1}{3} \left(H_{1} +H_{2} +H_{3} \right)\] 
Where  $H_{1} =\frac{\dot{A} }{A } $, $H_{2} =\frac{\dot{B} }{B } $ and $H_{3} =\frac{\dot{C} }{C } $
\begin{equation} \label{e24} 
	H=\frac{1}{3} \frac{k_{1} (n+2)}{n(k_{1} t+k_{2} )}  
\end{equation} 
The Scalar expansion ($\theta $) is
\[\theta =3H\] 
\begin{equation} \label{e25} 
	\theta =\frac{k_{1} (n+2)}{n(k_{1} t+k_{2} )}  
\end{equation} 
The deceleration parameter ($q$) is
\[q=\frac{d}{dt} \left(\frac{1}{H} \right)-1\] 
\begin{equation} \label{e26} 
	q=\frac{2(n-1)}{(n+2)} ,  
\end{equation} 
The Anisotropic parameter ($A_{m}$) is
\[A_{m} =\frac{1}{3} \sum _{i=1}^{3}\left(\frac{H_{i} -H}{H} \right)^{2}  \] 
\begin{equation} \label{e27} 
	A_{m} =2\left(\frac{n-1}{n+2} \right)^{2}  
\end{equation}
The Shear Scalar ($\sigma ^{2} $) is 
\[\sigma ^{2} =\frac{3}{2} A_{m} H^{2} \] 
\begin{equation} \label{e28} 
	\sigma ^{2} =\frac{1}{6} \left(\frac{k_{1} (n-1)}{n(k_{1} t+k_{2} )} \right)^{2}  
\end{equation} 

\section{Some Dynamical Properties }
The equation of state (EoS) for cosmic domain wall is
\begin{equation} \label{e29} 
	p=-\frac{2}{3} \rho  
\end{equation} 
From equation (\ref{e16}) and (\ref{e17}) we have  
\begin{equation} \label{e30} 
	\frac{\ddot{A} }{A } + \frac{\ddot{B} }{B } -\frac{\dot{A} \dot{B} }{A B } -\frac{\dot{B}^{2}}{B^{2}} =\kappa (p+\rho ) 
\end{equation}
Using equation (\ref{e29}), equation (\ref{e30}) yields 
The density ($\rho$) is
\begin{equation} \label{e31} 
	\rho =\frac{2}{\kappa n} \left(\frac{k_{1} }{(k_{1} t+k_{2} )} \right)^{2}  
\end{equation}
Form equation (\ref{e29}) we have 
The Pressure ($p$) is 
\begin{equation} \label{e32} 
	p=-\frac{4}{3\kappa n} \left(\frac{k_{1} }{(k_{1} t+k_{2} )} \right)^{2}  
\end{equation}
The cosmological constant ($\Lambda$) is
\begin{equation} \label{e33} 
	\Lambda =\frac{7}{3\kappa n} \left(\frac{k_{1} }{(k_{1} t+k_{2} )} \right)^{2}  
\end{equation} 
\section*{Jerk Paramater}
In cosmology, the term "jerk parameter" refers to a dimensionless quantity that characterizes the rate of change of the acceleration of the expansion of the Universe. This parameter plays a crucial role in understanding the dynamics of the cosmos and helps us investigate the nature of dark energy, a mysterious force driving the accelerated expansion of the Universe.\\
The expansion of the Universe was initially thought to be slowing down due to the gravitational pull of matter, which is the dominant component in the Universe. However, observations of distant supernovae in the late 1990s provided compelling evidence that the expansion is actually accelerating. This unexpected discovery led to the proposal of dark energy, a hypothetical form of energy that counteracts the attractive force of gravity, as the driving force behind this acceleration.\\
The jerk parameter, denoted as "$j$", is a higher-order cosmological parameter that comes into play when studying the evolution of the cosmic scale factor, $a(t)$, which describes how the Universe expands over time. It is related to the third derivative of the scale factor with respect to cosmic time, $a(t)$. is defined and discussed in the ref. (Blandford et al., 2004, Rapetti et al., 2007, Chiba and Nakamura, 1998, Visser, 2004, Luongo, 2013) \\
The cosmic scale factor is given by
\begin{equation} \label{e34} 
	a\left(t\right)=V^{1/3} 
\end{equation} 
Using equation (\ref{e23}) in equation (\ref{e34}) yields
\begin{equation} \label{e35} 
	a\left(t\right)={[n(k_1t+k_2)]}^{\frac{n+2}{3n}} 
\end{equation}
\begin{equation} \label{e36} 
	r=j\left(t\right)=\frac{1}{aH^3}\frac{d^3}{{dt}^3}\left(a\right) 
\end{equation}
From equation (\ref{e35}) and equation (\ref{e36}), we have
\begin{equation} \label{e37} 
	r=j\left(t\right)=\frac{2\left(1-n\right)\left(2-5n\right)}{{\left(n+2\right)}^2} 
\end{equation}
\section{Energy Conditions}
Energy conditions encompass a collection of principles and inequalities within the framework of general relativity. They serve as essential tools for characterizing the behavior of energy-momentum tensors in the fabric of spacetime. These conditions assume a pivotal role in our comprehension of the curvature of spacetime and its implications for extraordinary phenomena, including concepts such as wormholes or warp drives. In the following discussion, we will delve into an overview of several fundamental energy conditions.
\begin{enumerate}
	\item \textbf{Strong Energy Condition (SEC)}: The Strong Energy Condition posits that gravitational forces should always be attractive. In terms of the energy-momentum tensor, it is expressed as: $\rho + 3P \ge 0$
	\item \textbf{Null Energy Condition (NEC)}: The Null Energy Condition states that for any null vector $\mu^\alpha$, the following inequality must hold: 	$\rho + P \ge 0$
	\item \textbf{Weak Energy Condition (WEC)}: The Weak Energy Condition (WEC) asserts that the energy density observed by any observer must always remain non-negative, implying that it cannot be negative under any circumstances i.e. $\rho + P \ge 0$ and $\rho \ge 0$
	\item \textbf{Dominant Energy Condition (DEC)}: The Dominant Energy Condition (DEC) serves as a criterion that ensures the energy density observed by any observer must remain non-negative, implying that\\ $\rho \ge \mid P\mid$ i.e. $\rho- P\ge 0$ and $\rho+ P\ge 0$.
\end{enumerate}
from equations \ref{e31} and \ref{e32} we found that \\
\textbf{SEC} : 
\begin{equation} \label{e38}
	\rho + 3P = -\frac{2}{\kappa n} \left(\frac{k_{1} }{(k_{1} t+k_{2} )} \right)^{2}  \le 0
\end{equation}
\textbf{NEC}:
\begin{equation} \label{e39}
	\rho + P = \frac{2}{3\kappa n} \left(\frac{k_{1} }{(k_{1} t+k_{2} )} \right)^{2}  \ge 0
\end{equation}
and 
\begin{equation} \label{e40}
	\rho =\frac{2}{\kappa n} \left(\frac{k_{1} }{(k_{1} t+k_{2} )} \right)^{2} \ge 0
\end{equation}
\textbf{WEC}
\begin{equation} \label{e41}
	\rho + P = \frac{2}{3\kappa n} \left(\frac{k_{1} }{(k_{1} t+k_{2} )} \right)^{2}  \ge 0
\end{equation}
\textbf{DEC}
\begin{equation} \label{e42}
	\rho - P = \frac{10}{3\kappa n} \left(\frac{k_{1} }{(k_{1} t+k_{2} )} \right)^{2}  \ge 0
\end{equation}
\section{Statefinder Parameters}
The statefinder parameters is a set of dimensionless parameters introduced by Sahni \& Starobinsky (2000) to provide a more insightful understanding of the cosmic dynamics and the nature of dark energy. These parameters, denoted as ${r, s}$, are used to diagnose the evolution and characteristics of the Universe and its components. They are particularly useful for distinguishing between different cosmological models, such as the Cold Dark Matter with a Cosmological Constant ($\Lambda$CDM) model and the Standard Cold Dark Matter (SCDM) model, by identifying unique fixed points $(r, s)=(1,0)$ and $(r, s)=(1,1)$ respectively correspond to each model's properties. Statefinder parameters offer a geometric perspective on the Universe's expansion and are a valuable tool in cosmological studies.\\
The state finder parameter $r$ and $s$ are defined as follows:
\begin{equation} \label{e43}
	r=\frac{1}{aH^3}\frac{d^3}{{dt}^3}\left(a\right)  , \ \ \ s=\frac{r-1}{3(q-\frac{1}{2})}
\end{equation}
Using equations \ref{e26}, \ref{e35} and \ref{e37}, above equations yields
\begin{equation} \label{e44}
	\left\{r,s\right\}=\left\{\frac{2\left(1-n\right)\left(2-5n\right)}{{\left(n+2\right)}^2 },\frac{19n^2-32n+4}{9(n+2)(n-2) }\right\} 
\end{equation}
\begin{figure}[H]
	\begin{center}
		\centering
			\includegraphics[width=0.5\columnwidth]{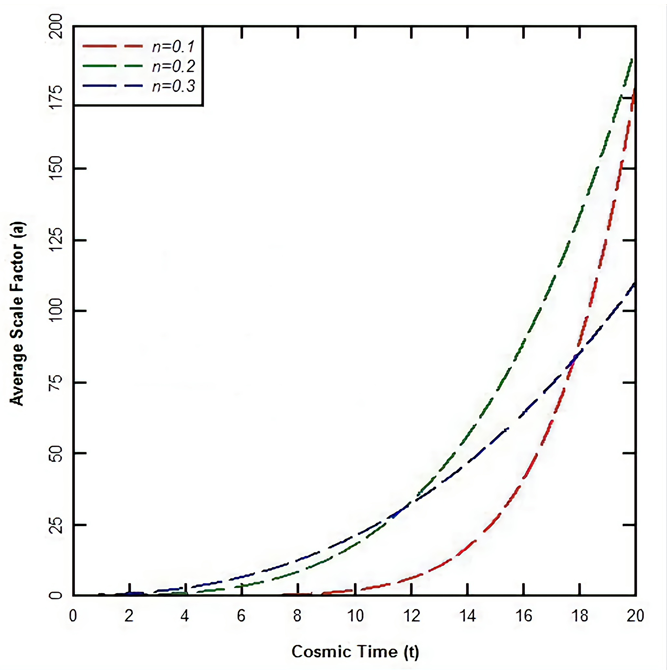}
		\caption[]{Variation of Average scale factor (a) against cosmic time (t) for $n=0.1,n=0.2,n=0.3$}\label{fig1}
	\end{center}
\end{figure}
\begin{figure}[H]
	\begin{center}
		\centering
			\includegraphics[width=0.5\columnwidth]{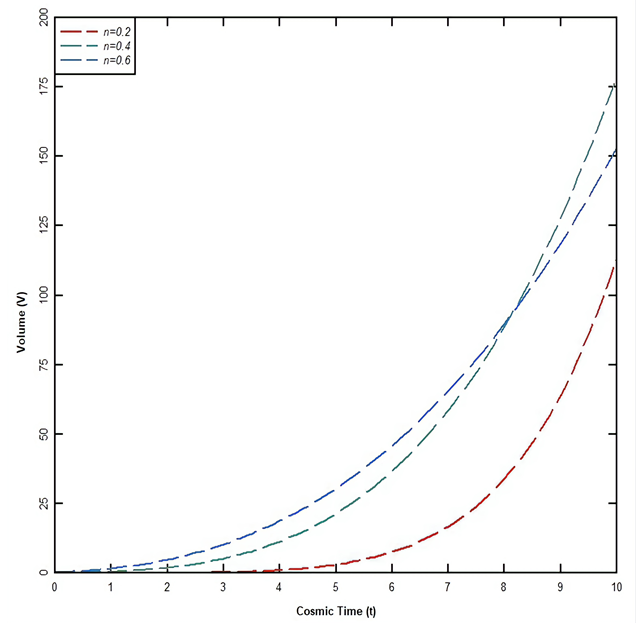}
		\caption[]{Variation of Volume (V) against cosmic time (t) for $n=0.2,n=0.4,n=0.6$}\label{fig2}
	\end{center}
\end{figure}
\begin{figure}[H]
	\begin{center}
		\centering
		\includegraphics[width=0.5\columnwidth]{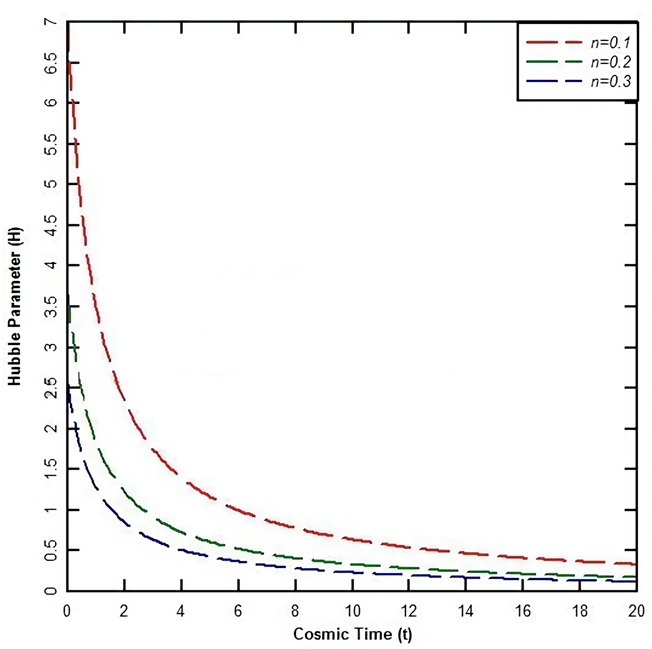}
		\caption[]{Variation of Hubble parameter (H) against cosmic time (t) for $n=0.1,n=0.2,n=0.3$}\label{fig3}
	\end{center}
\end{figure}
\begin{figure}[H]
	\begin{center}
		\centering
		\includegraphics[width=0.5\columnwidth]{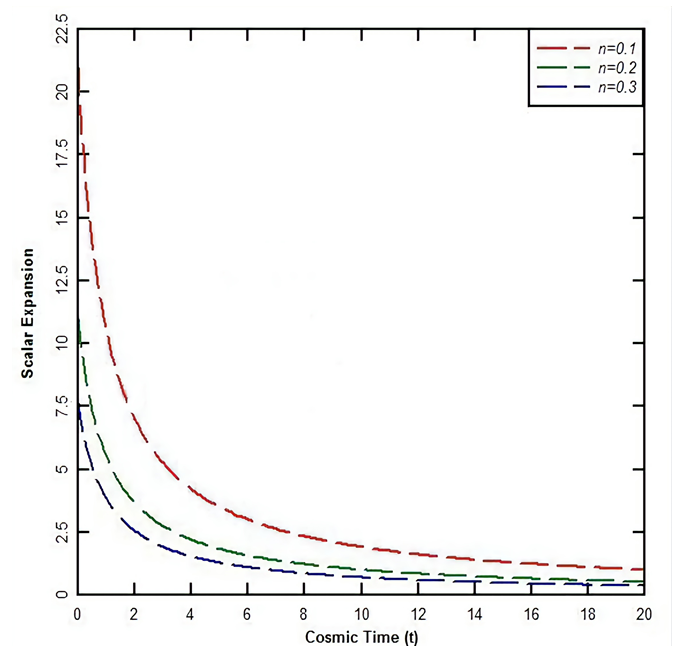}
		\caption[]{Variation of Scalar expansion ($\theta$) against cosmic time (t) for $n=0.1,n=0.2,n=0.3$}\label{fig4}
	\end{center}
\end{figure}
\begin{figure}[H]
	\begin{center}
		\centering
		\includegraphics[width=0.5\columnwidth]{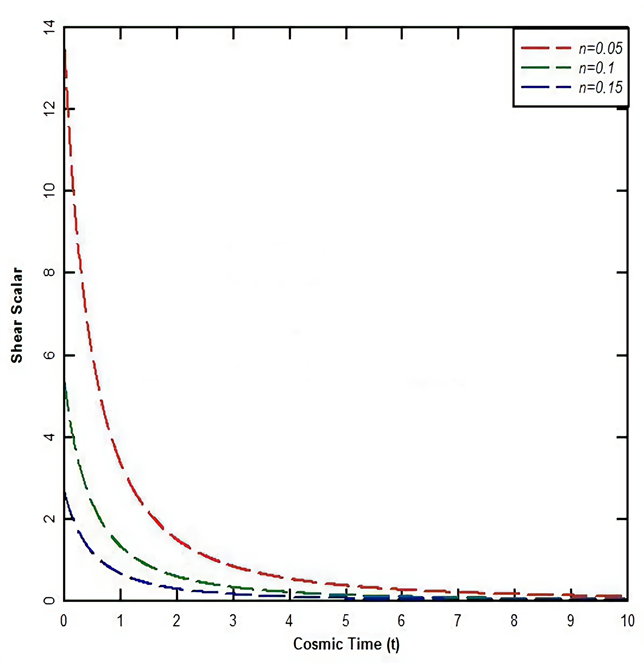}
		\caption[]{Variation of Shear scalar ($\sigma$) against cosmic time (t) for $n=0.05,n=0.1,n=0.15$}\label{fig5}
	\end{center}
\end{figure}
\begin{figure}[H]
	\begin{center}
		\centering
		\includegraphics[width=0.5\columnwidth]{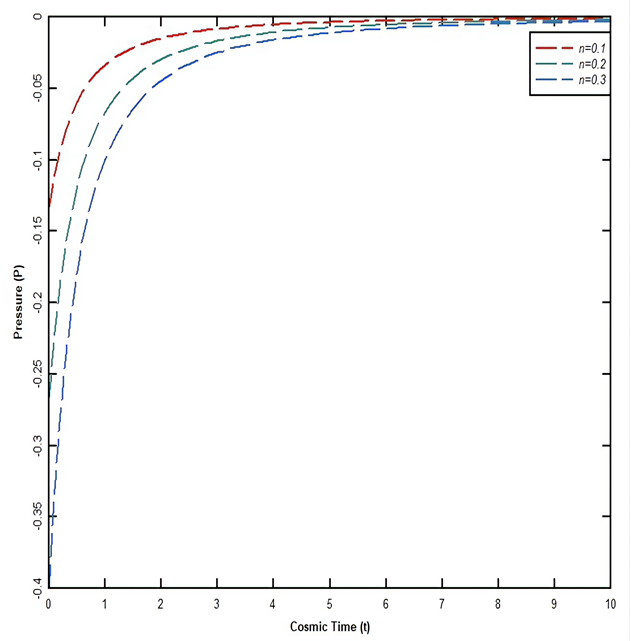}
		\caption[]{Variation of Pressure ($P$) against cosmic time (t) for $n=0.1,n=0.2,n=0.3$}\label{fig6}
	\end{center}
\end{figure}
\begin{figure}[H]
	\begin{center}
		\centering
		\includegraphics[width=0.5\columnwidth]{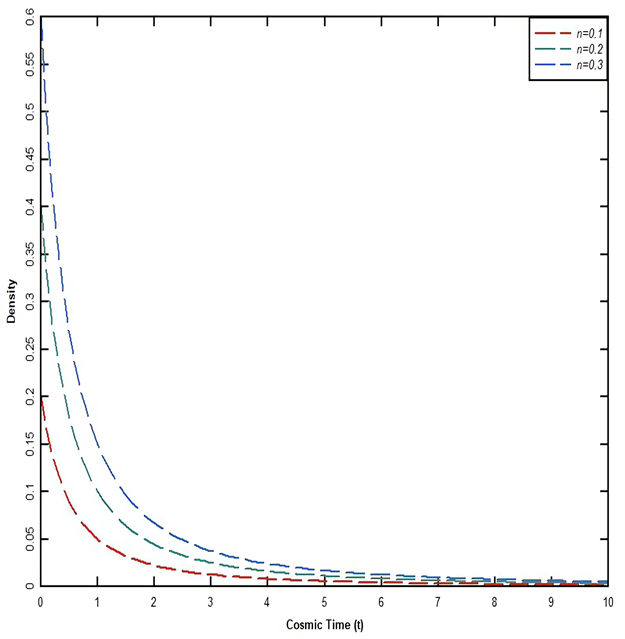}
		\caption[]{Variation of Density ($\rho$) against cosmic time (t) for $n=0.1,n=0.2,n=0.3$}\label{fig7}
	\end{center}
\end{figure}
\begin{figure}[H]
	\begin{center}
		\centering
		\includegraphics[width=0.5\columnwidth]{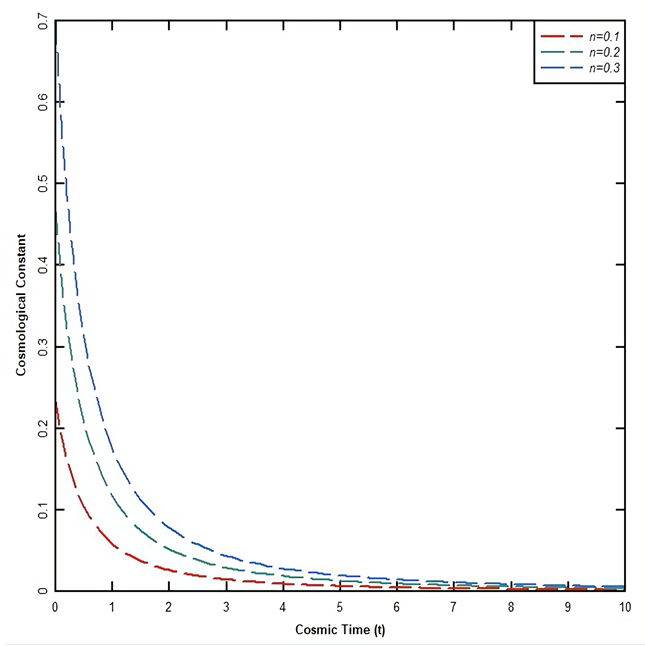}
		\caption[]{Variation of Cosmological constant ($\Lambda$) against cosmic time (t) for $n=0.1,n=0.2,n=0.3$}\label{fig8}
	\end{center}
\end{figure}
\begin{figure}[H]
	\begin{center}
		\centering
		\includegraphics[width=0.5\columnwidth]{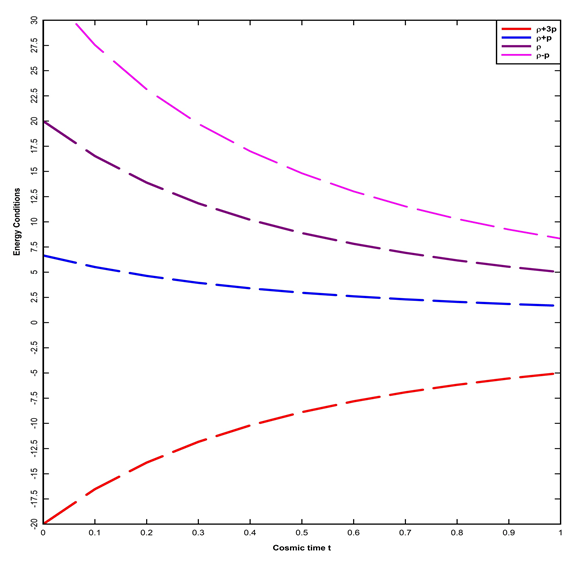}
		\caption[]{Variation of Energy Conditions (EC) against cosmic time (t) for $n=0.1$}\label{fig9}
	\end{center}
\end{figure}
\begin{figure}[H]
	\begin{center}
		\centering
		\includegraphics[width=0.5\columnwidth]{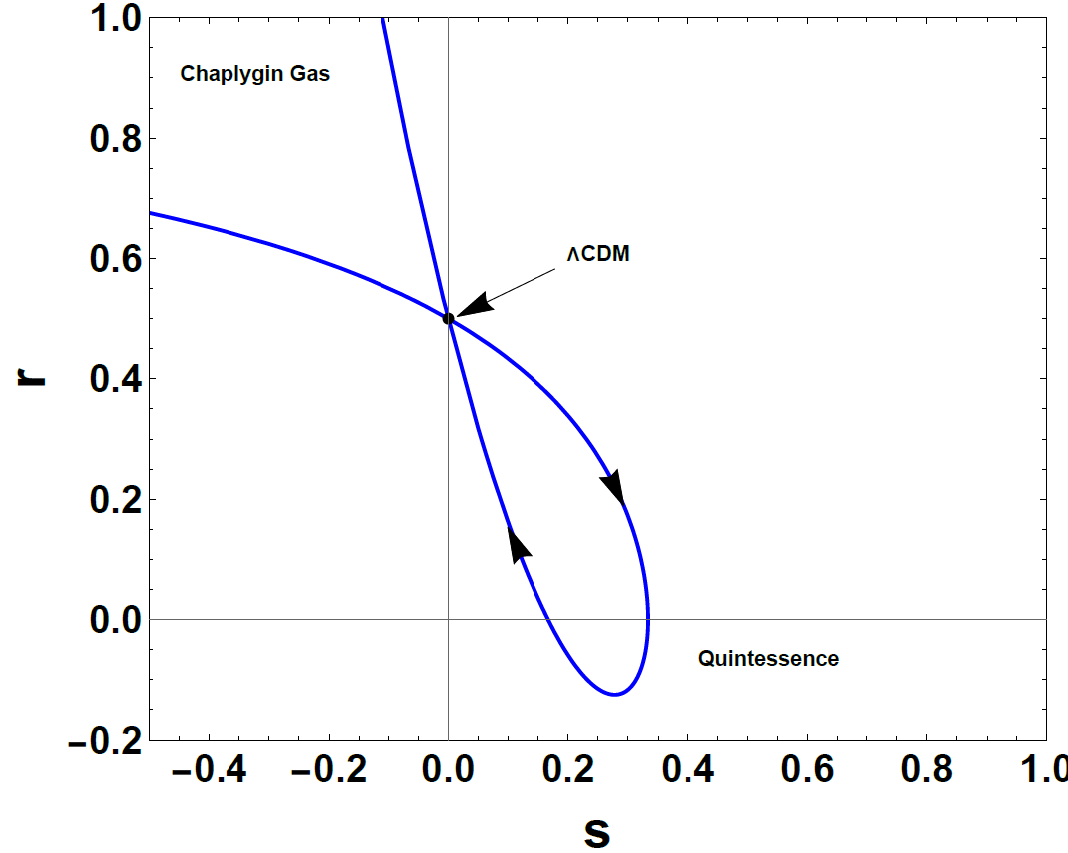}
		\caption[]{Evolution of $r-s$ trajectory}\label{fig10}
	\end{center}
\end{figure}
\section{Observations From Figures}
From the figure we have observed that,
\begin{itemize}
	\item The increasing graph of the average scale factor and volume of the Universe as depicted in Figure \ref*{fig1} and Figure \ref*{fig2} serve as compelling indicators of the Universe's expansion, a pivotal concept in our comprehension of the cosmos. These findings align with empirical observations like the redshift of light emanating from distant galaxies, providing strong support for the Big Bang theory and the notion of a Universe in a state of continuous expansion.
	\item It is observed that the Hubble parameter $H$ is decreasing function of cosmic time ($t$) in the positive region (Figure \ref*{fig3}). The positive value of $H$ confirmed about the expansion of the Universe. Initially, the rate of expansion is faster but later on it slows down as time increases and it will be zero for large time ($t$) (Figure \ref*{fig4}).
	\item We found that shear scalar is a diminishing function of cosmic time ($t$) (Figure \ref*{fig5}). The present model is not shear free except for $n=1$.
	\item The pressure vary from large negative value to small negative value (Figure \ref*{fig6}) and it tends to zero for large time $t$ ($t\rightarrow \infty$). A negative nature of the pressure shows that the existence of dark energy. Density is the decreasing function of cosmic time ($t$) (Figure \ref*{fig7}). It approaches to zero for infinite time.
	\item  We found the variable cosmological constant which is decreasing function of cosmic time ($t$) (Figure \ref*{fig8}).
	\item Through observations, it has been established that the Weak Energy Condition (WEC), Null Energy Condition (NEC), and Dominant Energy Condition (DEC) are all satisfied during the expansion of the Universe. However, it is noteworthy that the Strong Energy Condition (SEC) is found to be violated, as illustrated in Figure \ref*{fig9}.
	\item Figure \ref*{fig10} illustrates the progression trajectory of statefinder parameters corresponding to the specified $f(R,T)$ gravity model. 

\end{itemize}
Here all above quantities i.e. average scale factor $(a)$, volume $(V)$, Hubble parameter $(H)$, scalar expansion $(\theta)$, shear scalar $(\sigma)$, pressure $(P)$, density $(\rho)$ and cosmological constant $(\Lambda)$ are in arbitrary units. 

\section{Concluding Remark}
In the framework of the $\textit{f(R, T)}$ theory of gravity, we derived exact solutions to the field equations of an anisotropic Bianchi type-$VI_{0}$ cosmological model of the Universe with cosmic domain wall as an energy source.The metric potentials are finite at initial epoch, increasing as time increases and model does not have initial singularity. We have investigated the profile of cosmological and dynamical parameters in the framework of $f(R,T)$ gravity. In the present described model, we have used theoretical model as a cosmological constant for describing the nature of the dark energy and noticed that the cosmological constant is positive and dependent function of cosmic time ($t$). It indicates towards the observations of Supernovae Ia experiment (Riess, et al., 1998). We have found the existence of dark energy because of negative pressure. Also, we observed that the Universe is expanding as the Hubble parameter ($H$) is positive and the volume ($V$) of the Universe is increasing with cosmic time ($t$). Also, the rate of expansion of the Universe is decreasing with the cosmic time ($t$). It is good agreement with the observations of Pawar et al., 2021. The deceleration parameter (DP) has negative value for $0<n<1$ which pointing the accelerating phase of the Universe. From equation (\ref{e27}) and (\ref{e28}), it is clear that our Universe is verified as anisotropic and not shear free throughout the evolution of the Universe for ($n\ne 1$). Equation of state (EoS) plays major role to explore the nature of the Universe and we have used EoS for cosmic domain wall $\left(p=-\frac{2}{3} \rho \right)$ to describe the dynamical parameters of the model. We have observed the pressure and density are finite at initial time and decreasing functions of cosmic time ($t$). Also, we have explored the jerk parameter and compared with $\Lambda CDM$ model where the value of jerk is given by $j(t)=1$. In the present discussed model we found $j(t)\rightarrow1$ as $n\rightarrow0^+$ ($n>0$) from equation (\ref{e37}), which shows, present model approaches to $\Lambda CDM$ model. We found the cosmological constant $\Lambda$ is a function of cosmic time $t$ and agreed with the observations numerous researchers (Pop-lawski, 2006, Zel'dovich, 1968, Linde, 1974). We have observed the variable cosmological constant $\Lambda$ gives positive value, which indicates the expansion of  Universe is accelerating which reassemble with observations of present  the Universe's behavior. The statefinder parameter trajectory traverse through the $\Lambda$CDM fixed point and subsequently extend into the quintessence region $(r < 1, s > 0)$ before ultimately reaching the Chaplygin gas region $(r > 1, s < 0)$ as time unfolds into the distant future.
Finally, the Raychaudhuri equations demonstrate the satisfaction of the Null Energy Condition (NEC), Weak Energy Condition (WEC), and Dominant Energy Condition (DEC) signifies non-negative energy density, affirming the model's physical viability and the ongoing accelerated expansion of the universe. However, the Strong Energy Condition (SEC) is found to be violated. This SEC violation aligns with previous findings by (Sahoo et al.,2020, Patil et al. 2023), indicating an accelerated expansion of the Universe.

\end{document}